\documentclass[useAMS,usenatbib,a4paper]{mn2e}

\title[The origins of GMCs]
{Giant Molecular clouds: what are they made from, and how do they get there?}
\author[C. L. Dobbs, J. E Pringle \& A. Burkert]
{C. L. Dobbs\footnote{E-mail:
dobbs@astro.ex.ac.uk}$^{1,2,3}$, J. E. Pringle$^4$ and
A. Burkert$^{2,3}$\footnote{Max-Planck fellow}\\
$^1$ School of Physics, University of Exeter, Stocker Road, Exeter EX4
4QL \\
$^2$ Max-Planck-Institut f\"ur extraterrestrische Physik, Giessenbachstra\ss{}e, D-85748 Garching, Germany \\
$^3$ Universitats-Sternwarte M\"unchen, Scheinerstra\ss{}e 1, D-81679
M\"unchen, Germany \\
$^4$ Institute of Astronomy, Madingley Road, Cambridge, CB3 0HA }
\usepackage{amssymb}
\usepackage{amsmath}
\usepackage{graphicx}
\usepackage{epsfig}
\usepackage{multirow}
\usepackage{color}

\begin{document}
\date{\today}

\pagerange{\pageref{firstpage}--\pageref{lastpage}} \pubyear{0000}

\maketitle

\label{firstpage}
\begin{abstract}
We analyse the results of four simulations of isolated galaxies: two with a rigid spiral potential of fixed pattern speed, but with different degrees of star-formation induced feedback, one with an axisymmetric galactic potential and one with a `live' self-gravitating stellar component. Since we use a Lagrangian method we are able to select gas that lies within giant molecular clouds (GMCs) at a particular timeframe, and to then study the properties of this gas at earlier and later times. We find that gas which forms GMCs is {\em not} typical of the interstellar medium at least 50 Myr before the clouds form and reaches mean densities within an order of magnitude of mean cloud densities by  around 10 Myr before. The gas in GMCs takes at least 50 Myr to return to typical ISM gas after dispersal by stellar feedback, and in some cases the gas is never fully recycled. We also present a study of the two-dimensional, vertically-averaged velocity fields within the ISM. We show that the velocity fields corresponding to the shortest timescales (that is, those timescales closest to the immediate formation and dissipation of the clouds) can be readily understood in terms of the various cloud formation and dissipation mechanisms. Properties of the flow patterns can be used to distinguish the processes which drive converging flows (e.g.\ spiral shocks, supernovae) and thus molecular cloud formation, and we note that such properties may be detectable with future observations of nearby galaxies.
\end{abstract}

\section{Introduction}
Understanding how giant molecular clouds (GMCs) form is complicated by the
difficulty of observing, or determining the progenitors of molecular
clouds, whether they are small molecular clouds, HI clouds, or flows
of HI gas.\textcolor{white}{\let\thefootnote\relax\footnotetext{$\star$ E-mail:
dobbs@astro.ex.ac.uk}\footnotetext{$\dagger$ Max-Planck fellow}} A common scenario is to assume that gas is converted from HI
to molecular gas. In this case, the transitional stage may correspond
to HI self absorbtion (HISA)
\citep{Gibson2005,Strasser2007,Gibson2010}. Though HISA is ubiquitous
throughout the galaxy, it is not clear how it relates to molecular
cloud formation. Another possibility is that dense GMCs form from gas which
is molecular, but not forming stars \citep{Pringle2001}.
 In this case,
it may be possible to trace the
molecular gas with tracers such as HF and deuterium \citep{Monje2011,Pagani2011}.
The likelihood of finding such gas depends primarily on the UV field,
to determine what densities self shielding is effective, and how they
compare to the densities at which star formation takes place.
\citet{Pagani2011} propose that gas in nearby non-starforming dark
clouds cannot have been molecular for more than 6 million years. 
Alternatively it may even be objects which are
already forming stars, including IRDCs and
ordinary molecular clouds that are the main
precursors of GMCs.

A separate, but likely related question, is what is (are) the physical process(es) by which molecular clouds
form. GMCs may be formed by gravitational
instabilities, magnetic instabilities, thermal instabilities,
colliding flows or the
coalescence of smaller clouds. If GMCs form from the collisions,
or coalescence, of smaller clouds
\citep{Field1965,Tomisaka1984,Tomisaka1986,Kwan1987,Roberts1987}, 
then  these clouds must be either
denser HI or H$_2$. Numerous models 
 also hypothesise that molecular clouds form from colliding flows,
in which case the initial density of the gas is usually taken to be
sufficiently low that the gas is in the warm neutral (atomic) phase
\citep{Ball1999,Heitsch2006,
Vaz2007,Heitsch2008,Hennebelle2008,Banerjee2009,Ntormousi2011,Vaz2011}.     

Numerical simulations of galaxies have demonstrated the formation of GMCs from gravitational instabilities \citep{Shetty2006,Dobbs2008,Tasker2009,Dobbs2011,Tasker2011}
and cloud-cloud collisions \citep{Dobbs2006,Tasker2011,Dobbs2011}. Simulations of
colliding flows typically cover much smaller scales, e.g.\ 10's to 100's of
parsecs, and by necessity adopt fairly simplified initial conditions,
e.g.\ opposing flows in one direction, and a uniform initial
density. The colliding flows are assumed to arise from
stellar winds and/or supernovae \citep{Koyama2000,Heitsch2008,Ntormousi2011}, spiral shocks \citep{Leisawitz1982}, gravitational
instabilities, or turbulence in
the ISM \citep{Ball1999}. In most simulations the mechanism which is presumed to produce the
flows is not modelled, one exception being \citet{Ntormousi2011} who recently simulated two
supernovae whose expansion shells collide. 
To date though, the nature of colliding, or converging, flows has not been considered in a
galactic context. On slightly larger scales, simulations have also modelled a supernova driven ISM, where density enhancements occur at the intersection of large scale velocity flows \citep{deAvillez2004,Slyz2005,Joung2006,Dib2006,Gressel2008}. Again though, it is not clear whether, and on what size scales this picture applies to the ISM of observed galaxies, and whether it is reproduced in galactic models of the ISM. 

In this paper we analyse simulations of galaxies which
model GMC formation. We first investigate the density evolution of material
that makes up molecular clouds. We do not specifically differentiate between atomic and molecular gas, since our results are independent of the chemical nature of the ISM, and the boundary between atomic and molecular gas. 
However for a typical galaxy, low density may be a proxy for atomic gas, and likewise high density ($\gtrsim 10$ cm$^{-3}$) for molecular gas. We then show how the properties of velocity flows in our simulations indicate the physical processes which are controlling the dynamics of the ISM. The rest of the paper is divided as follows: in
Section 2 we provide basic details of the simulations and how we carried out
the analysis; in Sections 3--6 we present results for each model in
turn, with a brief discussion on some higher resolution simulations in Section 7; in Sections 8 and 9 we provide some discussion in the
context of theoretical models of the ISM, and our conclusions.

\section{Method}
\subsection{Calculations}
We use the results of four simulations for the analysis presented in
this paper. These simulations are i)  a galaxy with an imposed stellar 
potential, ii) a grand design spiral which uses
a imposed spiral potential, iii) as ii) but with a higher level of feedback,
and iv) a flocculent galaxy which includes a live stellar
component. We refer to these calculations as `No spiral', `Spiral
5\%', `Spiral 20\%' and `Flocculent' respectively. 
With the exception of the Flocculent model, these simulations have been presented in a
previous paper \citep{Dobbs2011}. 
In all the calculations we used a Smoothed Particle Hydrodynamics
(SPH) code.
We provide the main details of the
simulations below.
A frame from each
simulation is shown in Figure~1, at a time of 200 Myr.

The four simulations all model the ISM in an isolated galaxy, and
include self gravity, heating and cooling of the ISM \citep{Glover2007}, and 
supernovae feedback \citep{Dobbs2011}. In the first three simulations,
(described in detail in \citealt{Dobbs2011}), we use an external logarithmic potential to represent the gravity from stars and
a dark matter halo. In the Spiral 5\% and Spiral 20\% models we also
include a spiral component to the potential \citep{Cox2002}, which
produces an $m=4$ spiral density wave. The gas in these
simulations is placed in a 10 kpc radius disc, with a surface density
of 8 M$_{\odot}$ pc$^{-2}$. The mass per particle is 2500 M$_{\odot}$ (using 1 million particles), thus we can only consider GMCs of mass $\gtrsim 10^5$ M$_{\odot}$. We do however also discuss in Section~7 results from a couple of 8 million particle simulations, where we again used an imposed spiral potential.

In the Flocculent model, we set
up the galaxy similarly to the isolated galaxy shown in
\citet{Dobbs2010}, using the mkkd95 program, which is part of the NEMO \citep{Kuijken1995}. 
This is a publicly available program from the NEMO stellar dynamical software package (Teuben 1995).
 The galaxy contains
100,000 halo particles, 40,000 bulge particles, 1.1 million disc
particles.  Of the disc particles, 100,000 are star particles, and 1 million are gas.
The gas represents 4 \% of the mass of the disc, or
$2 \times 10^9$ M$_{\odot}$, so the mass per particle is 2000 M$_{\odot}$. Unlike the other calculations, which
exhibit a constant surface density, the surface density of the gas in
the Flocculent models falls off with radius.

In the first three models, we set a temperature threshold of 20 K,
though in the Flocculent model we increased this to 50 K. We insert
feedback, as described in \citet{Dobbs2011}. The supernovae feedback is included as thermal
and kinetic feedback, and is added when a number of conditions are
satisfied, but mainly that the density must exceed a critical value and
the gas must be gravitationally bound. We determine the number of
stars expected to form from the total mass of molecular gas in our
gravitationally bound region. We then multiply this by a star
formation efficiency parameter, $\epsilon$ and determine the number of
massive stars, and therefore energy to insert in the ISM. The energy
is distributed in the particles according to a snowplough solution.
We take a density criterion for inserting supernovae of 1000 cm$^{-3}$
in all the models except the flocculent galaxy, where we take 200 cm$^{-3}$.
For the Spiral 5 \%, No spiral and Flocculent
models, we take $\epsilon=0.05$, and for the Spiral 20 \% model we
take $\epsilon=0.2$.

As we give momentum to the particles when we insert supernovae feedback, problems associated with overcooling are reduced. Overcooling could at most correspond to a factor of two overestimation of the efficiency parameter $\epsilon$. However tests where we inserted purely kinetic energy, and half kinetic half thermal energy were not significantly different. Generally, the value of $\epsilon$ only provides a relative measure of the effectiveness of feedback in our calculations, its absolute value is subject to a large uncertainty reflecting not just the numerical implementation of feedback, but also for example the choice of initial mass function (IMF).

\begin{figure*}
\centerline{
\includegraphics[scale=1.2]{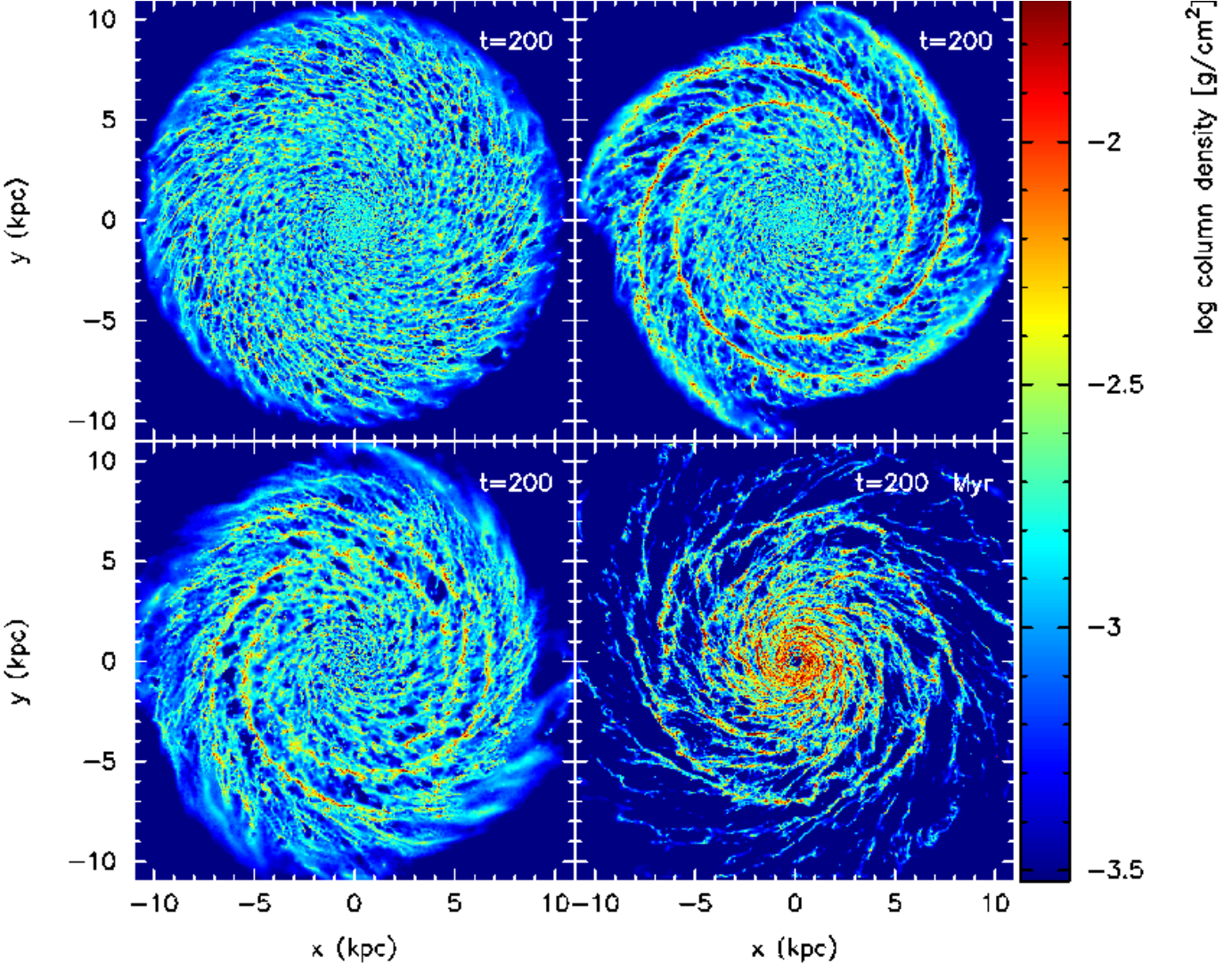}}
\caption{The gas column density is shown for the different
  calcualtions used in this paper, at a time of 200 Myr. The panels
  show a galaxy with a smooth potential (top left), a spiral potential
  (top right), a spiral potential with a strong level of feedback
  (lower left) and a flocculent galaxy with a live stellar component
  (lower right).}
\end{figure*}

\subsection{Analysis of gas density evolution and velocity flows}

\subsubsection{Tracing gas in GMCs}
Two of the key questions we wish to answer are: Is
the gas from which molecular clouds form typical or atypical ISM, and,
for how long is the ISM atypical? To do this we locate GMCs at a
particular time frame, then we trace the
particles in these clouds to earlier and to later times. 

We select clouds at a time $T_0$, (here 200 Myr), using a clump-finding algorithm (see
\citealt{Dobbs2008}). The algorithm selects cells above a given
surface density threshold criterion, and then labels all cells
which are adjacent as a single clump. We used cell sizes of 10 pc and 50 pc, and in both cases
adopted a density criterion of 50
M$_{\odot}$ pc$^{-2}$. For the cell size of 10 pc, the clouds were required to contain at least 30 particles. 
For the cell size of 50 pc, the clouds naturally contained more particles, with all the clouds in the spiral potential models containing at least 100 particles. The only disadvantage with taking cells of 50 pc was that there were few clouds for some models.

For the results we present in this paper, we show the case where we used a cell size of 10 pc.
 In Figure~2 we show the cumulative density
functions of this gas at times between $T_0-50$ Myr and $T_0$ for
the four different simulations. In Figure~3 we show the fraction of gas
above different densities versus time. In Figure~4 we study cloud
dispersal and show cumulative density
functions for times $T_0$ to $T_0+100$ Myr.    
Our results which used a cell size of 50 pc showed very little difference to Figures~2, 3 and 4. The only differences we noted were that there was slightly more noise, due to there being fewer clouds, and the lines on the panel for the flocculent galaxy in Figure~3 were slightly flattened. However we do not expect the results of our analysis to change. In this paper, we are considering the properties of gas that becomes especially dense. Our results are therefore primarily dependent on how the gas is organised in the galaxy, and thus the large scale gas flows, rather than the details of how the dense gas (i.e.\  GMCs) is selected.

\subsubsection{Tracing the velocity flows} 
Another question we wish to address is how does gas come together to
form molecular clouds? Thus we consider the velocity field of
the gas. For a 2D flow field ${\bf u}({\bf x},t)$, the resultant local
rate-of-strain tensor is
\begin{equation}
e_{ij} = \frac{1}{2} \left( \frac{\partial u_i}{\partial x_j} + \frac{\partial u_j}{\partial x_i} \right).
\end{equation}
This is a symmetric
tensor with two real eigenvalues $\lambda_1$ and $\lambda_2$ which
have dimensions of inverse time. The sum of the eigenvalues is then
the divergence of the flow,
\begin{equation}
\alpha = \lambda_1 + \lambda_2 = e_{ii} = \nabla \cdot {\bf u}.
\end{equation}
The continuity equation
\begin{equation}
\frac{D \Sigma}{Dt} = - \Sigma \nabla \cdot {\bf u},
\end{equation}
where $\Sigma$ is the surface density, and $D/Dt$ is the Lagrangian
derivative, then implies that the timescale on which (surface) density
is changing locally is given by $1/\alpha$. Positive $\alpha > 0$
implies expansion, and negative $\alpha < 0$ implies contraction. 
For example $\alpha=-0.25$ corresponds to gas that contracts on a timescale of 4 Myrs. We
show in Figure~5 contours of constant divergence (calculated in the
plane of the disc, thus neglecting vertical motions of the gas) overplotted on
column density for the different simulations. The divergence was
calculated across a grid of 100 pc spacing.

We also calculate the difference between the eigenvalues
\begin{equation}
\beta = |\lambda_1 - \lambda_2|,
\end{equation}
and in Figure~6 we plot each point in the flow in the $\alpha - \beta$ plane for
the simulations. We highlight the points in the $\alpha-\beta$ plane corresponding to high density gas in Figure~7. As we remarked above, points with $\alpha > 0$ represent expanding regions of fluid, whereas points with $\alpha < 0$ represent contracting regions. Points which best correspond to 1D colliding flows would lie in the region $\alpha \approx - \beta$. However this region is only a small subset of the $\alpha-\beta$ plane. 
The axis $\beta = 0$ represents those regions where the expansion/contraction is uniform in all directions, thus we might expect points in which gravity dominates the contraction to lie close to this line for negative $\alpha$. Conversely, points which exhibit negative $\alpha$, but $\beta>|\alpha|$ represent regions where there is convergence in one spatial dimension, but expansion in the perpendicular direction. 

\begin{figure*}
\centerline{
\includegraphics[scale=0.8]{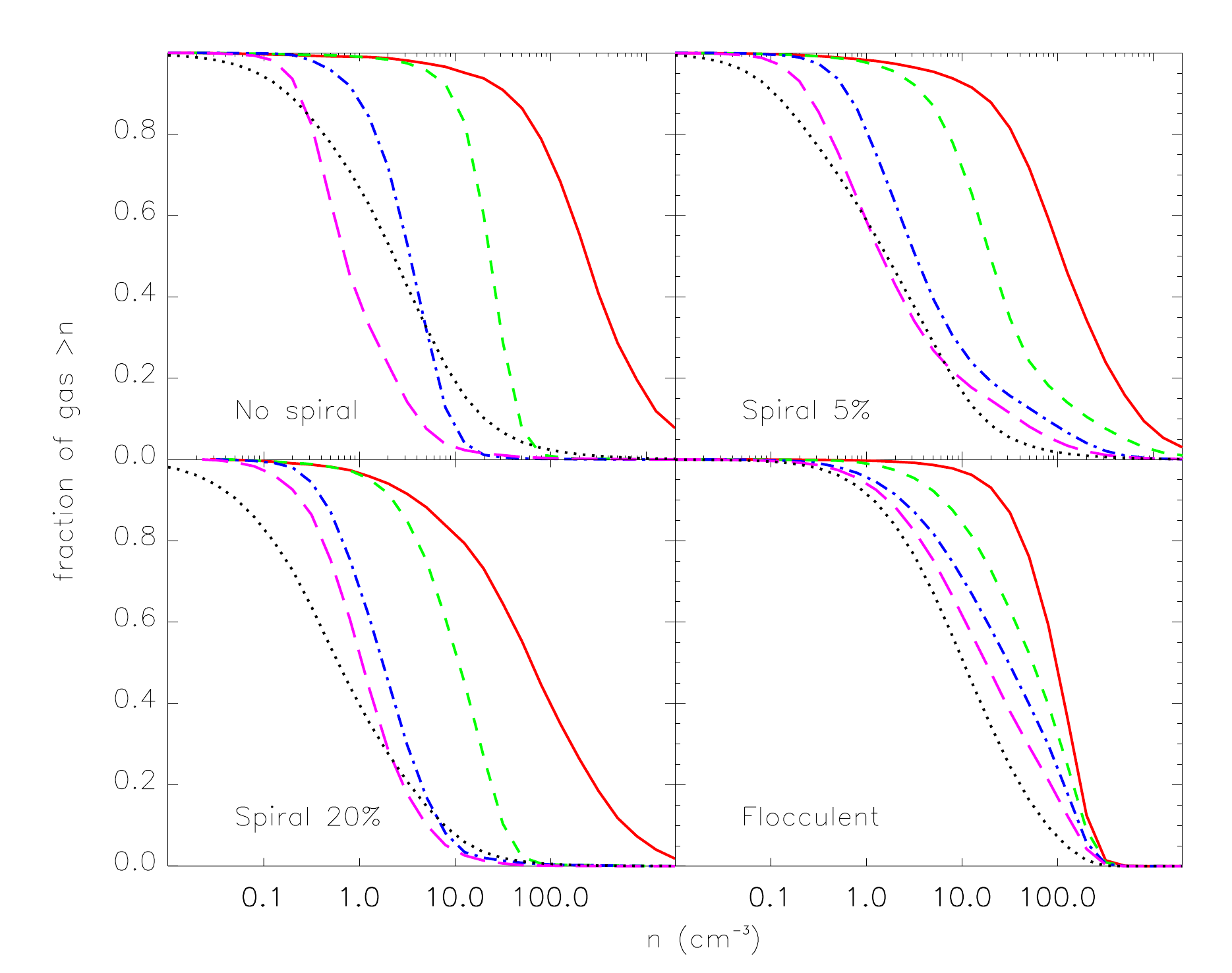}}
\caption{The cumulative distribution functions are shown for
  constituent gas of molecular clouds at times of $T_0-50$ (magenta, long dash),
  $T_0-30$ (blue, dot dash), $T_0-10$ (green, dash), $T_0$ (red, solid) Myr for the
  different models. The black (dotted) line shows the cdf for all the gas at a
  time of $T_0-50$ Myrs (note that the cdf for all the gas does not change significantly between $T_0-100$ and $T_0+100$ Myr). Even at $T_0-50$ Myr, the distribution of gas
  which forms the clouds is fairly atypical for the ISM of the galaxy. }
\end{figure*}

\begin{figure*}
\centerline{
\includegraphics[scale=0.8]{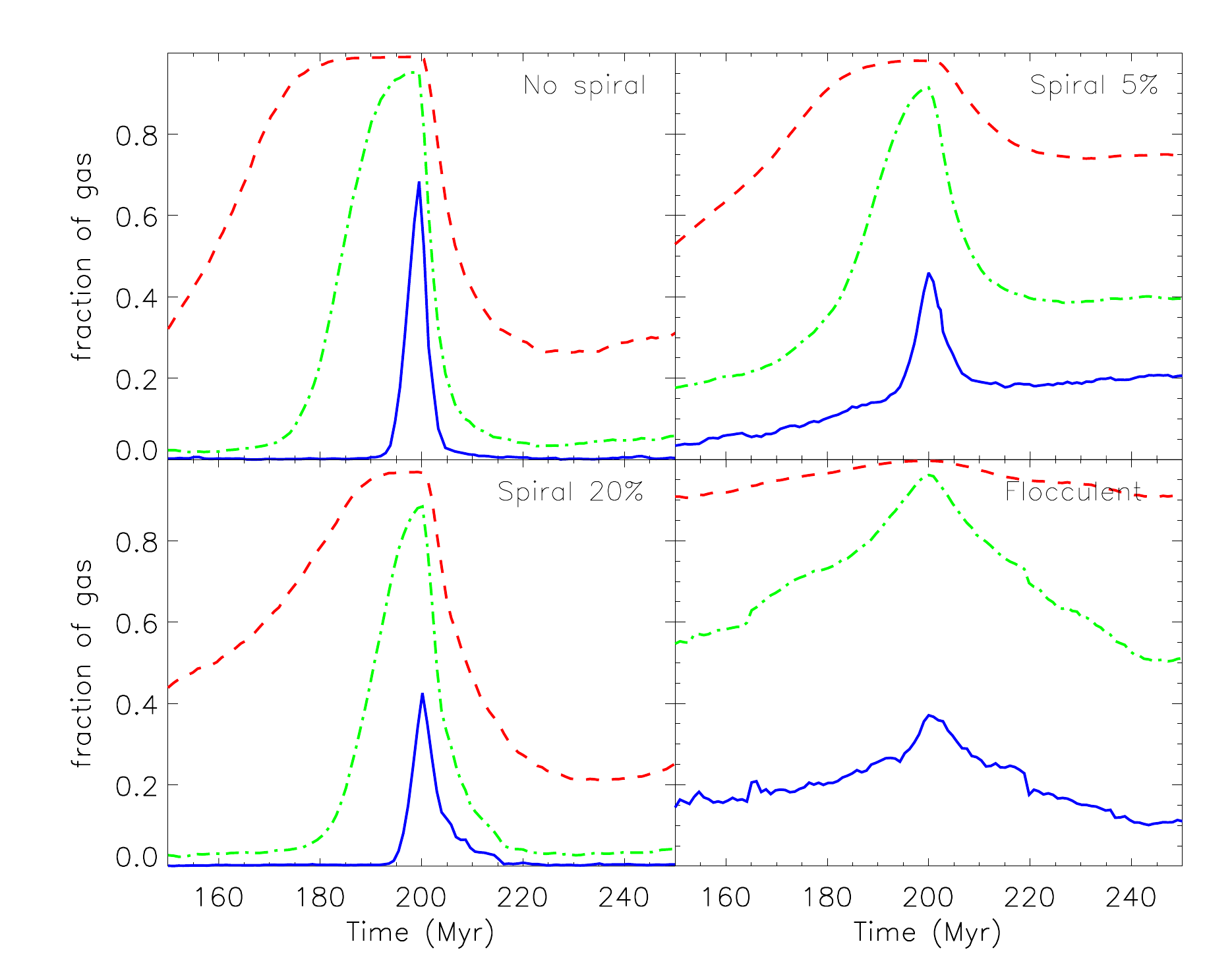}}
\caption{The fraction of gas over densities of 1 (red, dashed), 10 (green, dot dash) and
100 (blue, solid) cm$^{-3}$ in the molecular clouds is shown versus time for the different models.}
\end{figure*}

\begin{figure*}
\centerline{
\includegraphics[scale=0.8]{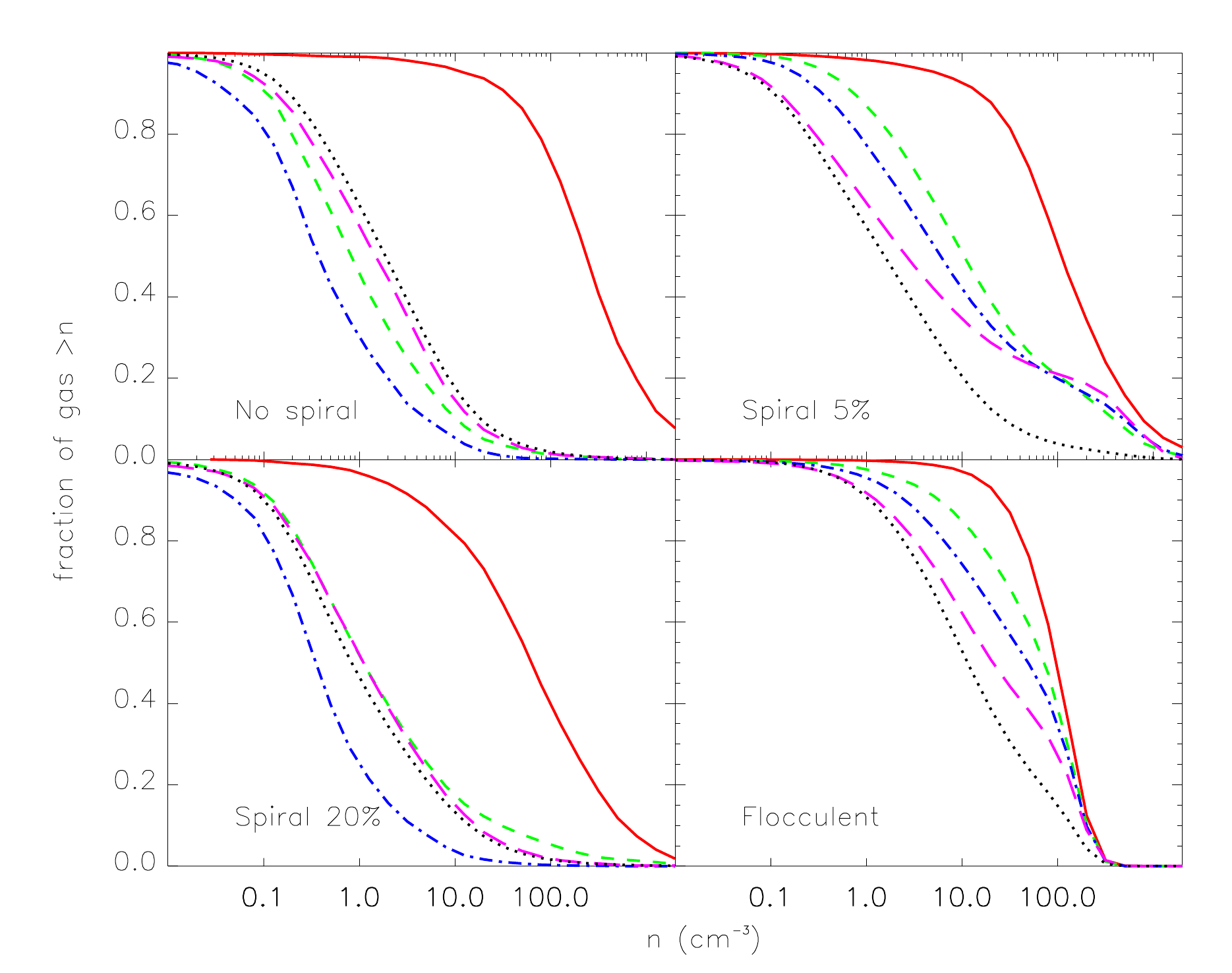}}
\caption{The cumulative distribution functions are shown for
  constituent gas of molecular clouds at times of $T_0$ (red, solid),
  $T_0+10$ (green, dash), $T_0+30$ (blue, dot dash), $T_0+100$ (magenta, long dash) Myr for the
  different models. The black (dotted) line shows the cdf for all the gas at a
  time of $T_0+100$ Myrs (note that the cdf for the total ISM at
  $T_0+100$ is very similar to that at $T_0-50$ shown in Figure~2). By
$T_0+100$ Myr, the gas in the clouds is indistinguishable from the ISM
for the No Spiral and Spiral 20 \% cases, but the gas is not fully
recylced for the models with a stronger spiral component.}
\end{figure*}

\begin{figure*}
\centerline{
\includegraphics[scale=0.8]{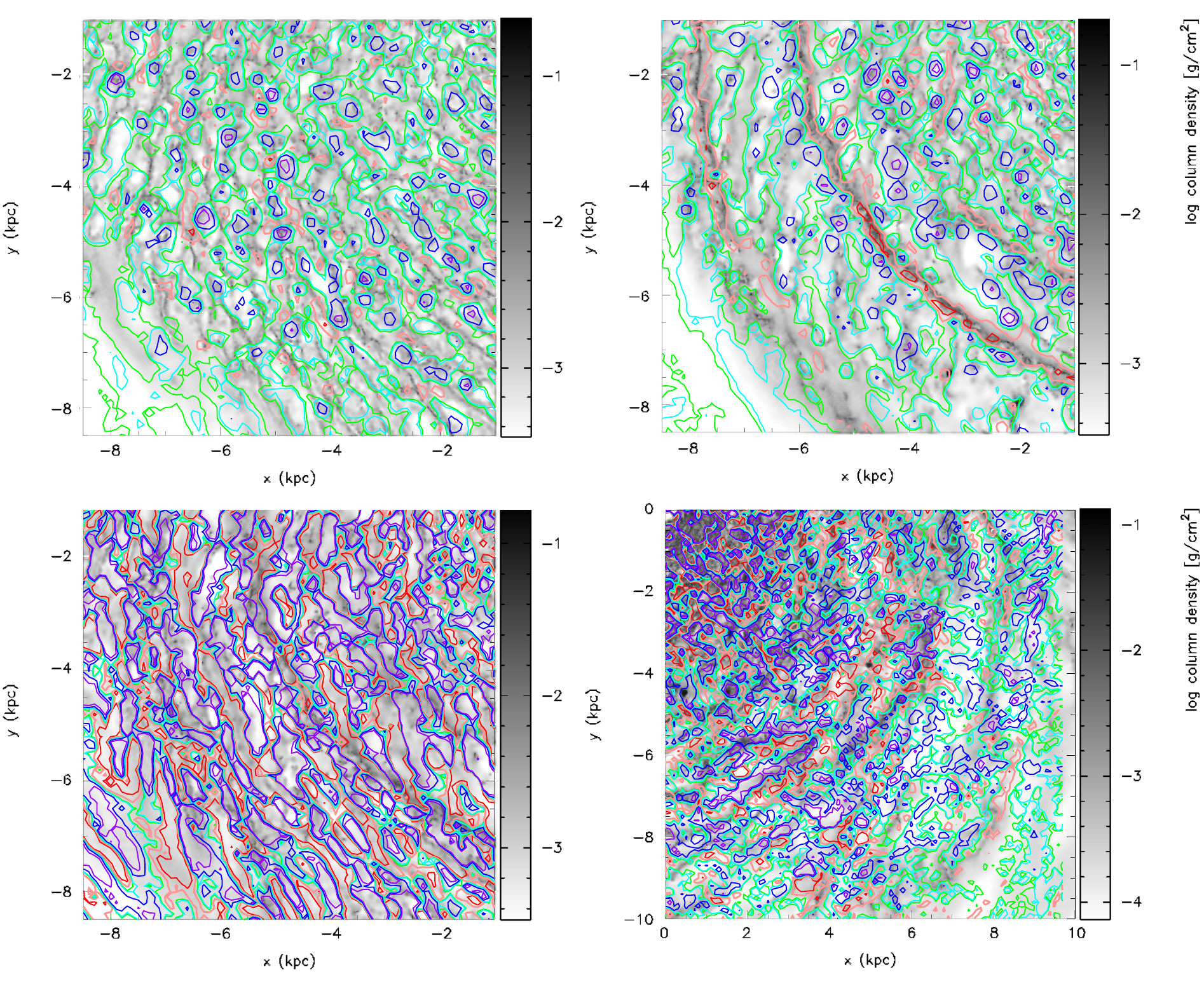}}
\caption{Contours of constant $\alpha$, where $\alpha$ is the
  divergence of the velocity field, are overplotted on the column
  density for the No spiral (top left), Spiral 5 \% (top right),
  Spiral 20 \% (lower left) and Flocculent (lower right) models. The
  contours show the timescales for gas to converge or diverge. The contours represent
  convergence on timescales of 4 (red), 10 (orange) and 100 (green) Myr and
  divergence on timescales of 4 (violet), 10 (blue) and 100 (cyan) Myr. In the
  spiral potential model with 5 \% feedback efficiency, the contours
  of strongest convergence clearly coincide with the spiral arms, but
  the patterns of convergence and divergence are much more random in
  the other models.}
\end{figure*}

\begin{figure*}
\centerline{
\includegraphics[scale=0.8]{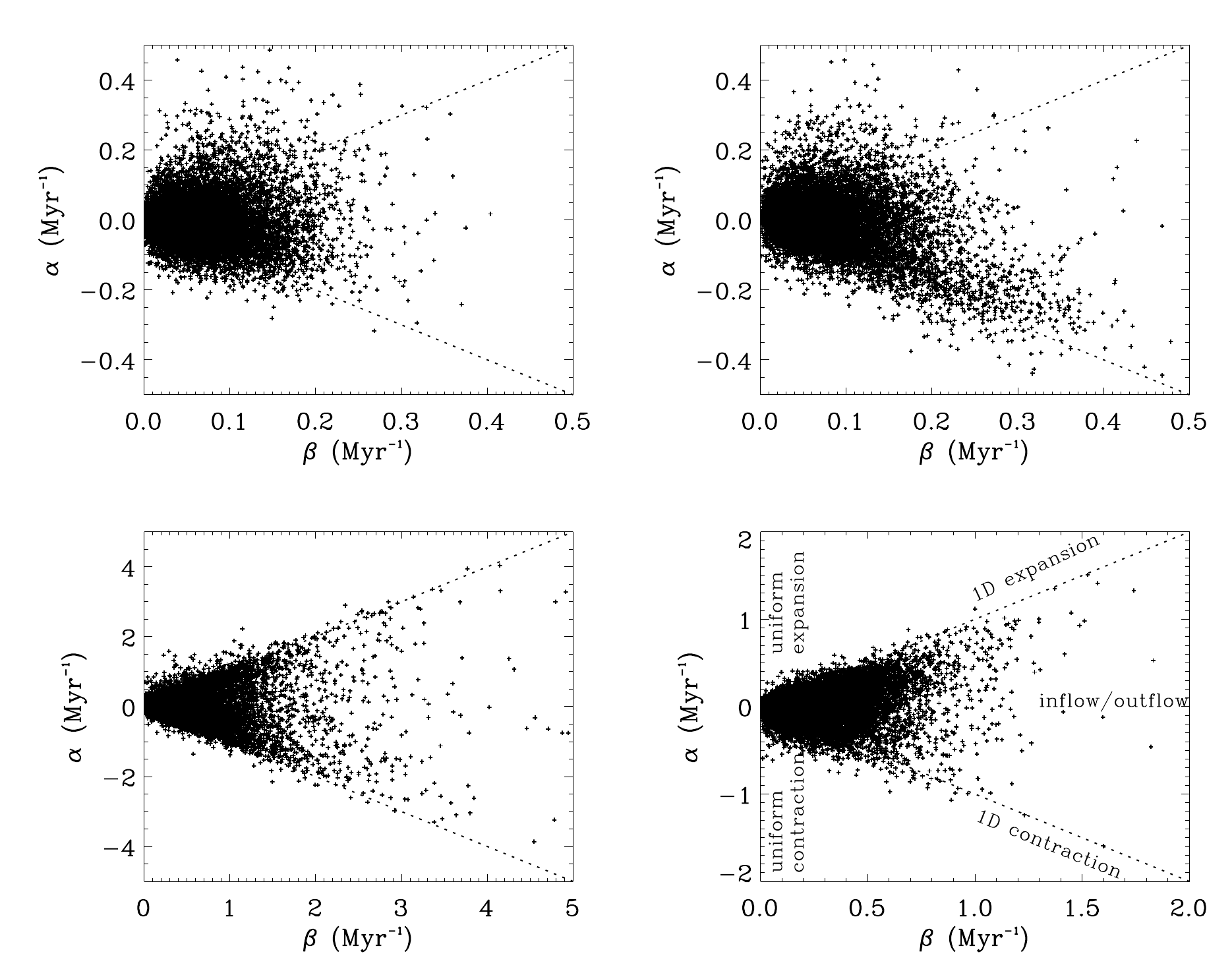}}
\caption{The values of $\alpha$ and $\beta$ are plotted for the
  No spiral  (top left), Spiral 5 \% (top right),
  Spiral 20 \% (lower left) and Flocculent (lower right)
  models. $\alpha$
  is the divergence of the
  flow ($\lambda_1+\lambda_2$), $\beta$ is the difference of the eigenvalues,
  $|\lambda_1-\lambda_2|$.   $\alpha$ and $\beta$ are calculated over a grid of points, with
  spacing 100 pc, for points between radii of 4 and 9.5 kpc. 
The top half of the plots correspond to expanding flows,
the lower half converging. The dotted lines mark lines where
$\alpha=\pm \beta$, corresponding to 1D contraction or expansion. The
various regimes for the flow are marked on the last panel (for the
flocculent model).}
\end{figure*}

\begin{figure*}
\centerline{
\includegraphics[scale=0.8]{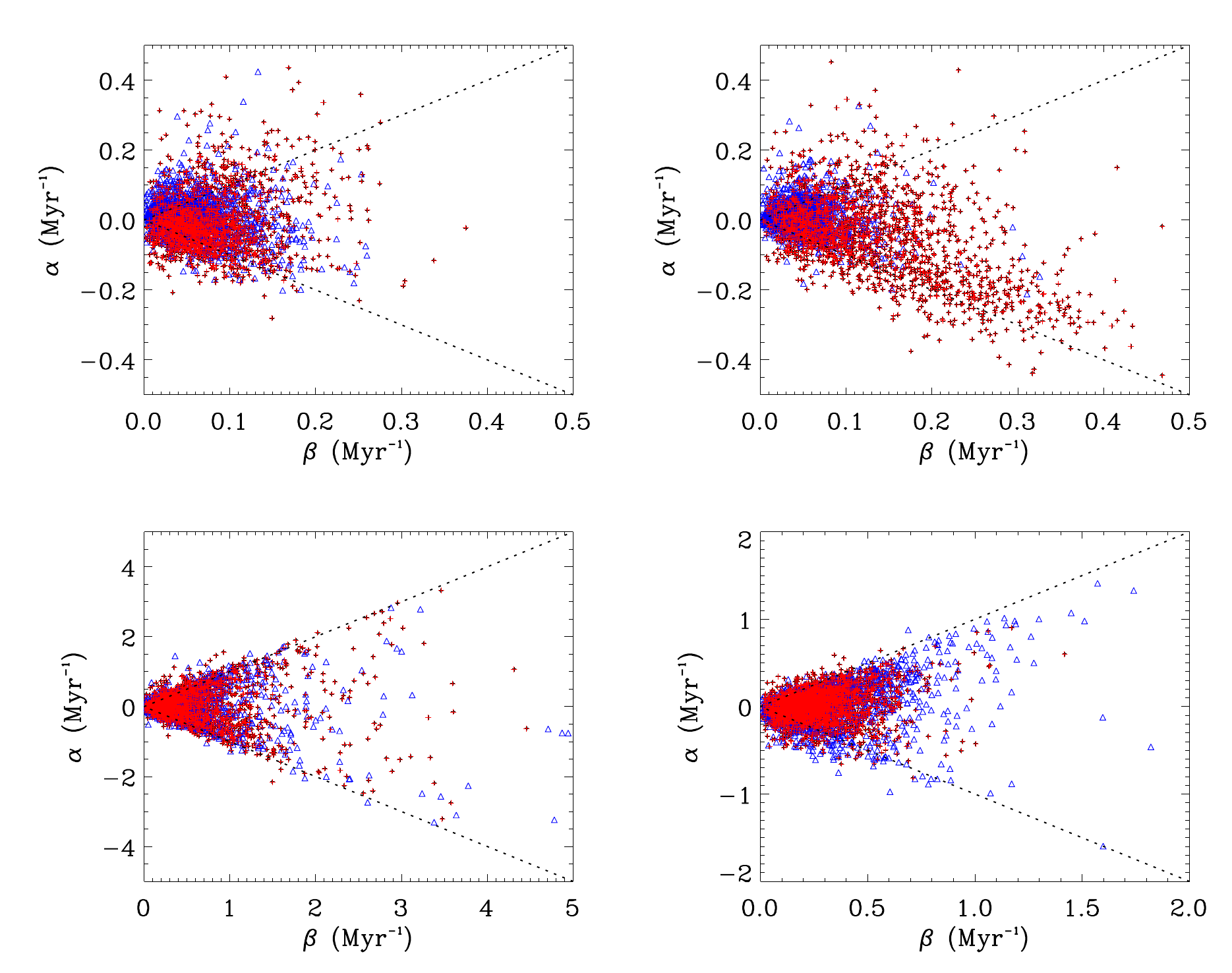}}
\caption{The values of $\alpha$ and $\beta$ are plotted as for Figure~6 except we only show the top 10\% most dense points (red crosses) and the 10\% least dense points (blue triangles).}
\end{figure*}

\section{No spiral, 5 \% efficiency feedback} 
\label{NoSp5}
Figure~1 (top left) shows the structure of the galaxy which uses a
stellar potential with no spiral component (`No Spiral'). The gas is dominated by structure on very small
scales, likely caused by thermal and gravitational instabilities, and
the stochastic nature of gas flows in the simulation. As noted in
\citet{Dobbs2011}, the structure is not very realistic compared to
observations, but this model provides a useful comparison. The structure and
evolution are also similar to other previous calculations
\citep{Wada2000,Tasker2009}.

\subsection{Density evolution}
In Figure~2 (top left) we show the density distribution of material which makes up the GMCs present at the time $T = T_0 =  200$ Myr. We plot here the cumulative density distribution, so that a point on each curve represents the fraction of gas which has densities greater than $n$.  The black (dotted) curve represents the density distribution of all of the ISM at time $T_0-50$ Myr. 

We note first that even at time $T_0 - 50$ Myr the density distribution of material which is going to end up in clouds at time $T_0$ is already considerably different to the ISM distribution. The median density of the gas is lower and the density distribution is more uniform with 80\% of the gas having densities in the range 0.5 cm$^{-3} < n < 5$ cm$^{-3}$, covering one order of magnitude, whereas the typical ISM has 80\% in the range 0.3 cm$^{-3} < n < 30$ cm$^{-3}$, covering two orders of magnitude. As we approach time $T_0$ the gas density distribution remains similarly narrow and steadily
increases in density. The median value increases from $n \approx 0.8$ cm$^{-3}$ to
$n \approx 20$ cm$^{-3}$ over a timescale of 40 Myr. In the final 10 Myr,
however, the evolution proceeds more rapidly. The median density
increases by an order of magnitude from $n \approx 20$ cm$^{-3}$ to $n
\approx 200$ cm$^{-3}$ and the density distribution becomes much less uniform,
with a larger high density tail. 

In Figure~3 (top left) we plot the time dependence of the fractions of gas above various densities which is in GMCs at time $T_0$. The median density of the ISM is approximately $\bar{n} \approx 2$ cm$^{-3}$. At time $T_0 - 30$ Myr essentially all of the gas has density $n > 1$, but none of it has $n>10$. From this Figure it is evident that the timescale for the appearance of the highest density gas (with $n > 100$ cm$^{-3}$) is much shorter than the timescale on which the intermediate density ($n> 10$ cm$^{-3}$) and lower density gas ($n > 1$ cm$^{-3}$) appears. Indeed the highest density gas only starts to appear in the last 5 Myr. 

From Figure~3 (top left) it is evident from the asymmetries of the
various curves that once the clouds form and feedback ensues the
density decreases rapidly. This is also seen in Figure~4 (top left) where we plot the cumulative density distributions at later times $T \ge T_0$ post feedback. Within 10 to 20 Myr the density {\em distribution} has decreased by around {\em three} orders of magnitude. Eventually, here shown at time $T = T_0 + 100$ Myr, the density distribution of the material that made up the clouds at time $T_0$ is indistinguishable from that of the ISM as a whole. Thus the material which makes up the clouds is fully mixed back into the ISM. 

\subsection{Velocity fields}
In Figure~5 we plot contours of constant $\alpha$ superposed on the
underlying density distribution at time $T_0 = 200$ Myr. Here the
asymmetry between contraction and expansion timescales is 
apparent. There are almost no contours corresponding to
contraction on a timescale of 4 Myr, corresponding to $\alpha=-0.25$, whereas there are many
corresponding to expansion on that timescale. Similarly in Figure~6
there are almost no points with $\alpha <- 0.25$ whereas there are many
with $\alpha > 0.25$. 
From Figure 7 we also see that there are many dense points with $\alpha<0$ and $|\alpha|<\beta$, indicating gravitational collapse. There are also a few regions which are converging (see contours in Figure~5) where the density is low, presumably where non gravitational effects, e.g.\ thermal instabilities and stochastic flows, act on the gas. However the fraction of points with $\alpha<0$ and $|\alpha|<\beta$ is higher compared to the other calculations. Thus self gravity has a greater influence on the gas flows in this simulation.  
\section{Imposed spiral pattern, 5\% feedback efficiency}
\label{Sp5}

In this model the feedback is the same as in Section~3.1 but a spiral
pattern is included in the applied stellar potential.
This $m=4$ spiral pattern is clearly apparent in the gas column density
(Figure~1, top right).
\subsection{Density Evolution}

In Figure~2 (top right) we show the density distribution of that material
which ends up in GMCs at time $T_0 = 200$ Myr for the
model with a spiral potential. In contrast
to the case with no superposed spiral, the distribution at time $T_0
- 50$ Myr is less different from the underlying ISM, with 80 \% of the
gas with 3 cm$^{-3}<n<300$ cm$^{-3}$, spanning two orders of magnitude. Prior to cloud formation
the gas tends to exhibit a wider range of densities compared to the
case without a spiral potential. By $T_0 -
30$ Myr however the material which will make up the clouds is distinctly
atypical. The median density ($\bar{n} \approx 3$ cm$^{-3}$) is about 2--3 times
larger than the typical ISM. At $T_0 - 10$ Myr, the median density has
increased to $\bar{n} \approx 20$ cm$^{-3}$ and 80\% of the material has
densities in the range 3 cm$^{-3}< n < 200$ cm$^{-3}$. 

From Figure~4 we see that in this case the amount of energy and momentum input as a result of star formation and feedback is not sufficient to disperse all of the dense clouds. Thus even some 100 Myr later, although the low density end of the density distribution resembles that of the ISM, some 20\% of the gas that was in clouds at time $T_0$ still has densities above $n \ge 100$ cm$^{-3}$.
Thus the long-term properties of the ISM have not yet reached an equilibrium.

\subsection{Velocity Field}

In Figure~5 (top right), in contrast to the case considered in Section~\ref{NoSp5}, the contours of $\alpha < -0.1$, that is regions contracting on timescales $\tau < 10$ My are strongly correlated with the density enhancements which we would identify as spiral arms. Regions expanding on similar timescales (that is, contours of $\alpha > 0.1$) can be found not only just downstream of the main density enhancements, but also spread throughout the interarm regions. 

From Figure~6 (top right) we see that the nature of the converging
flow is here quite different to the model with no spiral
potential. Here asymmetry between converging and diverging flows is
more pronounced. The converging flows with $\alpha < - 0.1$, that is
timescales $\tau < 10$ Myr, are scattered roughly around the line
$|\alpha| \approx \beta$, indicating that the convergence is mainly
along one direction. Figure~7 again indicates that these points are 
predominantly regions of higher density, and that the spiral shocks preferentially compress gas which is already relatively dense and cold.
Flows diverging on timescales
$\alpha > 0.1$ fall mainly to the left of the $\alpha = \beta$ line in
Figure~6 indicating that such flows are diverging in all directions.

\section{Imposed spiral pattern, 20\% feedback efficiency}
\label{Sp20}

In this model the effects of feedback are so strong that the presence
of the underlying spiral pattern is almost obliterated (Figure~1 lower
left). 
Thus this model lies intermediate between the model in Section~\ref{NoSp5} which has no imposed spiral pattern and the model in Section~\ref{Sp5} where because of the weaker feedback  the imposed spiral pattern has a much stronger influence on the structure.

\subsection{Density evolution}

From Figure~2 (lower left) we find again that at time $T = T_0 - 50$ Myr
the distribution of densities in material that will make up the GMCs
at time $T_0$ differs from that of the average ISM. 
The distribution of densities is
more uniform than that in the mean ISM, but is not as uniform as that
found in the gas in Section~\ref{NoSp5}. The density distribution increases by a factor of about 10 
between $T = T_0 - 50$ Myr and $T = T_0 - 10$ Myr. In the final
10 Myr the densities increase most rapidly, until at $T_0$ they span 5 cm$^{-3} < n < 500$ cm$^{-3}$.  After star formation
and feedback occur the density of the gas that made up the clouds
decreases rapidly. After 10 Myr the densities decrease by two orders
of magnitude and the density distribution is essentially the same as that of the average ISM. But the material is clearly still expanding because at $T = T_0 + 30$ Myr the densities decrease further by factors of 2 -- 3. As for the case discussed in Section~\ref{NoSp5} at time $T = T_0 + 100$ Myr, the density distribution of the material that was in the GMCs is indistinguishable from that of typical ISM material. 

\subsection{Velocity fields}

For this case it is evident from Figure~5 (lower left) that the timescales on which
both convergence and divergence occur are much shorter than in the
case with lower feedback efficiency (Section~\ref{Sp5}). This is also
evident in Figure~6  (lower left) where we have needed to change the scales for both
$\alpha$ and $\beta$ by a factor of 10. We also find that the regions
showing most rapid convergence (contours of $\alpha < 1$, or
timescales $\tau < 1$ Myr) are not as strongly correlated with density
as in the case with weaker feedback (Section~\ref{Sp5}). Indeed, and as also seen in Figure~7, some
rapidly contracting regions seem to have very low density, although it
is noticeable that these are almost always to be found next to regions
of rapid expansion ($\alpha > 1$), as expected if the compression is induced by stellar feedback rather than gravity. 
Comparing Figure~5 (lower and
upper left) it is evident that the underlying spiral potential has a
much weaker effect on the dynamics of the ISM in this case.
As the distribution of points in the $\alpha-\beta$ plane is independent of the gas density (Figure~7), the stellar feedback appears to be controlling the flows in the ISM regardless of density enhancements in the gas or stars.

Figure~6  (lower left) shows that all converging flows with
timescales $\tau < 1$ Myr (that is, $\alpha <-1$) have $\beta \ge
|\alpha|$. This implies that while these flows have strong
convergence in one direction, they also  display significant
divergence in the perpendicular direction -- like a stream of water
striking a brick wall. For similar timescales ($\tau < 1$ Myr),
flows that display net expansion have $\alpha > 1$. Although there are some
points with $\alpha < \beta$ in the majority of cases are characterised by
$\beta \ge \alpha$. This is somewhat surprising as one might imagine
that strong feedback might tend to generate expansion in all
directions, whereas the evidence in Figure~6  suggests that the strong net expansion is accompanied typically by contraction along one axis. 
 
\section{Flocculent galaxy, 5 \% efficiency feedback} 
In this model, spiral structure is generated by perturbations
in the stellar component, so the gas is not as structured as in the No Spiral
model, but neither is there a grand design structure like the Spiral 5\%
model. For the flocculent galaxy, the gas surface density decreases
with radius. Thus gas near the centre tends to be high surface
density regardless of whether it lies in a cloud. Hence for this model
we only select clouds in a torus between $r=3.5$ and $5.5$ kpc, where
the average gas surface density is similar to the other models.

\subsection{Density evolution}
From Figure~2 (lower right), we again see that at time $T_0-50$ Myr, the
distribution of densities differs from that of the average ISM. The distribution of
material is denser than the typical ISM. At $T_0-50$ Myr the median density is
around $n=10$ cm$^{-3}$ compared to $\bar{n}\sim$ 8 cm$^{-3}$ for the typical ISM. 
The median density increases to $50$ cm$^{-3}$ at $T_0-10$
Myr. There is a slightly more marked increase in density the last 10 Myrs
until $T_0$, although there is a lower critical density for supernovae
feedback in this calculation compared to the others. 

The evolution of the fraction of gas above given densities is quite
different from the other simulations.  There is a much higher fraction
of gas above 10 and 100 cm$^{-3}$ for the whole 100 Myr period shown
in Figure~3 (lower right). In Figure~4 (lower right), we see that even after 100 Myr the material
in the clouds is denser than the typical ISM. This could be because,
like in the model with a spiral potential and 5~\% efficiency, clouds
are not disrupted by feedback. So we followed the evolution of
some of the clouds in this model. We found that gas is not confined to
regions the size of molecular clouds (i.e.\ 10's or 100's parsecs), rather the gas is much more
widely spatially distributed prior to and after $T_0$. However the gas does
remain in spiral arms which are not
disrupted. As one perturbation dissipates, the gas retains its density
and then becomes part of another spiral arm. Something like this is
shown in Figure~3 of \citet{DB2008}. 

We checked whether increasing the level of stellar feedback altered the density evolution in the flocculent model. With stronger ($\epsilon=10$ \%) feedback, the gas is fully recycled, presumably the feedback is more effective in transferring dense arm material to low densities, above the plane of the disc, or in the interarm regions. The fraction of gas versus time for the clouds (Figure~3) also shows narrower peaks, and very dense ($>100$ cm$^{-3}$) gas only occurs around $T_0$. Though gas with densities $>1$ cm$^{-3}$ is still overabundant compared to the other calculations.

\subsection{Velocity fields}
The velocity field looks somewhere in between the cases with a fixed
spiral potential, and the feedback dominated case (Sections 4
and 5). Some regions of converging flows are coincident with
dense spiral arms, but others are not. The distribution of $\alpha$
and $\beta$ shows some similar characteristics to the Spiral 20\%
case, though the timescales are not as short. Some of the dense gas (Figure~7) exhibits 
$|\alpha|>\beta$ and $\alpha<0$, indicating more uniform convergence, though similar to the Spiral 20\% case, other regions of convergence are accompanied by outflows in the perpendicular direction.

\section{Higher resolution models with a spiral potential}
Whilst we do not have the means to rerun all our simulations at higher resolution, we did apply our analysis to a couple of higher resolution simulations, which most resemble the 5 \% efficiency case with the imposed spiral potential (Section~4). We investigated the GMCs in a simulation presented by \citet{DGCK2008}, which uses 8 million particles (model 2) with a mass resolution of 312 M$_{\odot}$ per particle. This simulation did not include self gravity or feedback, rather just the spiral potential and the thermodynamics. However the behaviour was very similar to that of the 5 \% efficiency case with the imposed spiral potential (Section~4), for example the range of densities span three orders of magnitude prior to and after GMC formation. This is not so surprising since, as we discuss in the next sections, the dynamics are driven foremost by the spiral potential. The only difference is the hump apparent at higher densities in  Figure~4 (top right) does not occur, as in the absence of self gravity the GMCs are always disrupted by shear, and after 100 Myr the gas is fully recycled.

Finally we also examined an unpublished calculation, which uses an $m=2$ spiral potential, 8 million particles (again with a mass resolution of 312 M$_{\odot}$), which includes self gravity and feedback. Again the properties of the gas before and after GMC formation are similar to the model with 5 \% efficiency and a spiral potential (see Figure~8). For both the higher resolution simulations, the gas is atypical ISM at least 30 Myr prior to and after GMC formation.
\begin{figure}
\centerline{
\includegraphics[scale=0.55, bb=50 0 400 330]{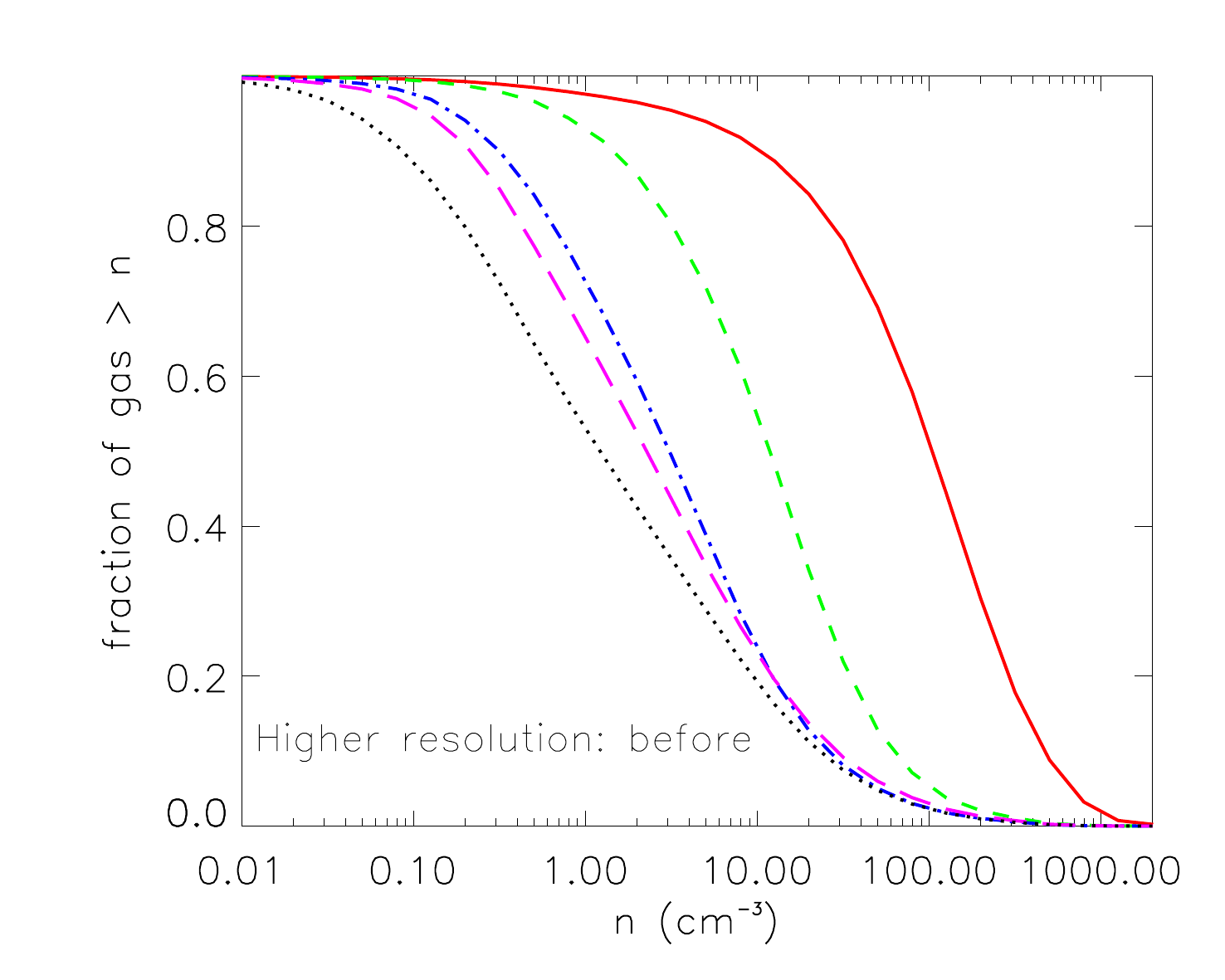}}
\centerline{
\includegraphics[scale=0.55, bb=50 0 400 310]{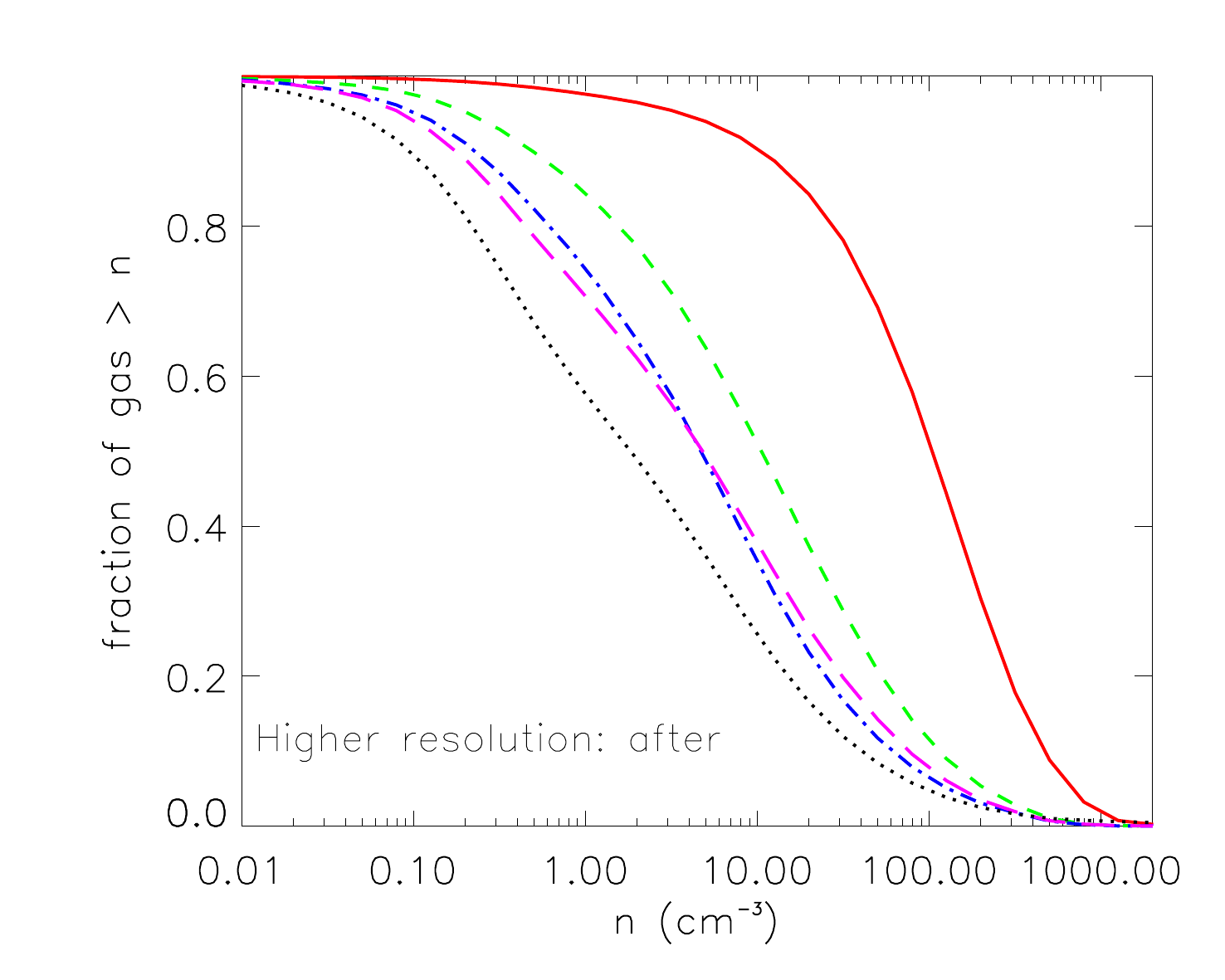}}
\caption{The cumulative distribution functions are shown for
  gas from a simulation with an imposed spiral potential and 8 million particles. The upper and lower panels show the constituent gas from GMCs selected at 200 Myr, at earlier and later times. The colour / line types for the upper and lower panels are the same as Figures 2 and 4 respectively.}
\end{figure}

\section{Comparison of models with theories of the ISM}
To make a molecular cloud it is necessary to gather a sufficient quantity of gas from the interstellar medium into a small enough volume. The questions we have attempted to address in this paper are: 

\begin{enumerate}

\item which parts of the ISM are gathered together, and

\item what mechanism(s) enable this to happen?

\end{enumerate}

We stress again that in this paper we have concentrated on only massive ($> 10^5$ M$_{\odot}$) GMCs. Although lower mass clouds may form in similar ways, we cannot conclude anything about lower mass clouds from our results. We also note that we have not included magnetic fields in our analysis.

\subsection{From what gas do molecular clouds form $\ldots$}

Many models of molecular cloud formation start with an ISM which is uniform in density, with a density appropriate to the mean density of the ISM as a whole (typically $\bar{n} \sim 1$ cm$^{-3}$;  e.g.\ \citealt{Elmegreen1991,KOS2002,Kim2006,Heitsch2006}) and  then investigate instabilities of various kinds (thermal, gravitational, magneto-dynamic). In contrast, \citet{Pringle2001} argued that it is difficult to form GMCs from such low density material on the necessary timescales (of order $\sim 5 - 10$ Myr), and that the gas from which GMCs form must already be dense -- within an order of magnitude of the mean cloud density, and much denser than the mean for the ISM. \citet{Pringle2001} then made the (tacit) assumption that GMCs form from typical ISM material and concluded that a large fraction of the ISM must consist of dense, possibly molecular, gas.

From our simulations it is evident that neither of these scenarios is fully correct. In each of our simulations the material which forms GMCs is significantly denser than the mean ISM by 10 Myr before cloud formation takes place (Figure~2). Thus the simple argument made by \citet{Pringle2001} is validated, and models which try to form GMCs on short timescales from gas with low average ISM densities are not likely to succeed. However, the tacit assumption made by \citet{Pringle2001} that clouds form from typical ISM material is not sustained.  Indeed (Figure~2) at least 30 Myr, and in some cases  50 Myr, before cloud formation takes place, the density distribution of the material destined to be part of GMCs differs significantly from the average for that of the ISM as a whole. 

\subsection{$\ldots$ and which mechanism, or mechanisms, gather it together? }

The interstellar gas in galaxies is characterised by a highly inhomogeneous, turbulent and multiphase medium. A number of mechanisms give rise to this of which the dominant ones are:

\begin{enumerate}

\item {\em Thermal effects}. It has long been noted (e.g.\ \citealt{McKee1977, Burkert2000}) that the ISM is a multiphase medium and that this is at least in part due to the complexity and multi-valued nature of the cooling functions. Such a structure is likely to be non-steady. As gas enters the cool phase it becomes denser and so contracts, and as it enters a warm or hot phase it expands. Thus such a structure of necessity gives rise to internal flows (\citealt{Elphick1991,Elphick1992}; see the discussion in \citet{Pringle2007}, Chap. 8). 

\item {\em Feedback. } \citet{McKee1977} postulated that the multiphase nature of the ISM is regulated predominantly by supernova explosions, and indeed it is such events that we have in mind when we insert star-formation induced feedback into our simulations.

\item {\em Self-gravity.} Although it is has been established \citep{DBP2006} that self-gravity is {\em not} necessary for the formation of GMCs, and although it is apparent that many, if not most, GMCs are not self-gravitating as a whole \citep{Heyer2009,Dobbs2011a} it is clear that self-gravity must play a role at least in the denser parts of the GMCs in which star formation is occurring, and also helps form more massive GMCs \citep{Kwan1987,Dobbs2008}. 

\item {\em Non-axisymmetries in the underlying stellar potential.} If the underlying stellar potential is non-axisymmetric, non-circular perturbations will be induced in the velocity field. These can give rise to shocks, especially in the higher density, cooler component of the ISM. We consider two cases (i) a steady spiral potential with fixed pattern speed, to mimic global tidally induced spiral structure (e.g. M~51) and, (ii) a model in which the underlying stellar component is gravitationally unstable, to model flocculent galaxies ( cf. \citealt{Wada2011}). 

\end{enumerate}

We have analysed the local, vertically-averaged velocity flows in our simulations by considering the eigenvalues, $\lambda_1, \lambda_2$ of the two-dimensional rate of strain tensor, $e_{ij}$, on a size scale of 100 pc (Figures~5 and~6). Plots of the divergence $\alpha = \lambda_1 + \lambda_2 =  \nabla \cdot {\bf u}$ versus $\beta = | \lambda_1 - \lambda_2|$ are given in Figure~6. In this Figure each point represents the nature of the flow locally at a position in the model. The area of this Figure in which points lie at larger values of $|\alpha|$ (shorter timescales) tells us about the nature of the flows which are contracting ($\alpha < 0$) or expanding ($\alpha > 0$) the most rapidly. As discussed already, points which best approximate 1D colliding flows (often adopted as initial conditions for molecular cloud formation, e.g.\ \citealt{Heitsch2006,Banerjee2009,Vaz2011}) exhibit $\alpha \approx - \beta$. Unsurprisingly, all our models contain some points which satisfy this criterion. However only the Spiral 5\% model preferentially contains flows which are of this nature.

For the No Spiral case (Section~3), the random motions in the ISM are induced by a combination of thermal effects, feedback and self-gravity. We find that the rapidly contracting flows (timescales $< 4$ Myr) do so more uniformly (i.e. have more points for which both $\lambda_1 < 0$ and $\lambda_2 < 0$) than for the other cases, especially for the denser gas (Figure~7). This seems to imply that in this case self-gravity plays a strong, but not always a dominant, role in the final gathering of material as it forms a GMC. 

As mentioned above, for the case of a superimposed spiral with moderate feedback (Spiral 5\%, Section 4), the rapidly contracting flows lie preferentially along the line $\alpha \approx - \beta$ indicating that the final contraction occurs predominantly in the one-dimensional manner. This is not a surprise since the shock induced by the spiral potential will tend to produce a one-dimensional compressional flow. We have remarked above that the cloud formation process differs here from that usually assumed because the density of the material is much higher than the mean ISM density. It is worth noting here that it is the cooler, denser component (say, molecular) of the ISM which undergoes the stronger shock and so undergoes the greater degree of compression (Figure~7). This is evidenced by the fact that the high density gas has much larger eigenvalues $(|\lambda_1|, |\lambda_2|)$ in Figure 7 than the low density gas. 

In the same superimposed spiral potential, but with much stronger feedback (Spiral 20\%, Section~5), the feedback is sufficiently energetic that the non-axisymmetric effects of the potential are to some extent overwhelmed (Figures~1 and 5).  If, in Figure~6, we compare this case with the Spiral 5\% case, we note that the flow timescales are here much shorter. This must imply that in this case the internal motions of the ISM are dominated by feedback, as anticipated by \citet{McKee1977} and modelled by many authors \citep{Rosen1995,deAvillez2004,Slyz2005,Joung2006}. The points in Figure~6 lie almost entirely within the region $\beta \ge |\alpha|$, implying that the local flows are highly non-uniform, with a large velocity in one direction (expansion or compression) together with a smaller, but usually non-negligible, velocity {\em of the opposite sign} in the perpendicular direction (similar to the colliding flows models of \citet{Audit2005} which employ open boundary conditions). The lack of the strong correlation between rapidly converging flows (regions of large, negative $\alpha$ in Figure~5) and the local density (confirmed by the fact that the low and high density points in Figure~7 are similarly distributed in the $(\alpha,\beta)-$plane) implies that it is indeed strong feedback which is the main driver.

The behaviour of the ISM subject to a `live' stellar disc (`Flocculent', Section 6), in which the stellar arms are non-steady, forming and then being wound and stretched by galactic shear, has been modelled by a number of authors \citep{Li2005b,DB2007,Robertson2008,Dobbs2009,Hopkins2011,Wada2011}. The local stellar densities in the arms change typically on an orbital timescale (say, $\sim 100$ Myr) and there is a tendency for the arms to bifurcate and/or to merge with other arms. In Figures~5 and~6 we see that the timescales on which the local gas density changes is often much shorter than this, e.g.\ $\tau>0.2$ corresponding to a timescale of $<5$ Myr. Here again (Figure~6) most, but not all,  of the points have $|\alpha| \le \beta$, so again often contraction in one direction is accompanied by expansion in the other (through feedback and/or shear). 

\section{Conclusions}

We have investigated models for the formation of molecular clouds which encompass most of the physical mechanisms usually envisaged, including predominantly self-gravity (No Spiral), compressive shocks in a spiral arm (Spiral 5\%), feedback from supernovae (Spiral 20\%) and flow subject to non-steady potential minima in a `live' stellar disc (Flocculent). All these mechanisms are capable of forming molecular clouds and it is of interest to try to determine which (if any) dominates in a particular galaxy. A related question is whether the timescales for cloud formation and dispersal, and the nature of the ISM which goes into making up clouds, differ in these different scenarios.

Whilst we generally see a transition from predominantly low to high densities when GMCs form (e.g.\ from atomic to molecular gas), this transition occurs over a long (10's of Myrs) time period. Thus our main finding is that GMCs do not form from typical or average interstellar material. At 30 Myr, and in some cases even 50 Myr before cloud formation, the density distribution of the cloud-forming material differs significantly from that of the ISM as a whole. 
Within 10 Myr prior to cloud formation the mean density of gas destined to make up a cloud is within an order of magnitude of mean density of the cloud.
Given the strongly inhomogeneous nature of the ISM, whatever the mechanism which gathers material together -- whether it depends strongly on density (e.g. self-gravity) or not (e.g. supernova explosions) -- it is that material which is already of higher density that is more likely to form the GMCs \citep{Pringle2001}.

We find that in most cases the effect of feedback is to fully recycle material to the ISM (Figure~4). 
The exceptions to this are the self-gravitating long-lived entities which form in Spiral 5\% model, and the clouds in the spiral arms of the Flocculent model. Both of these simulations are characterised by a clear spiral structure and a lower level of feedback. In these models, it appears  (Figure~4) that material which forms GMCs has a tendency to remain denser than average. This raises the possibility that the abundances of gas to be found within star-forming regions (i.e.\ dense clouds and flocculent spiral arms) might differ significantly from the abundances of gas in the ambient ISM.

We also find (Figures~5, 6 and 7) that the nature of small-scale flows within the ISM can give strong clues as to the nature of the dominant mechanism by which GMCs are formed in a particular galaxy. For example a comparison of the distributions of points in Figures~6 and 7 for the Spiral 5\%  and the Flocculent models, both of which have similar levels of feedback, indicates a significant difference between models of spiral galaxies in which the gas flows relative to the arm pattern (as in density wave theory) or in which the pattern and circular flow essentially co-rotate (see also \citealt{Wada2011}). Likewise, a comparison between the models which form GMCs predominantly due to self-gravity of the gas (No Spiral) and which form GMCs predominantly due to the self-gravity of the underlying stellar component (Flocculent), both of which present a `flocculent' appearance, shows that the groupings of points within the $(\alpha, \beta)$--plane are easily distinguishable. Lastly, our model where the gas flows are dominated by stellar feedback (Spiral 20\%) exhibits different characteristics again compared to the other simulations. With future facilities, such as ALMA, such comparisons can become reality.

\section{Acknowledgments}
We thank an anonymous referee for a helpful report which improved the paper. 
CLD acknowledges funding from the European Research Council for the
FP7 ERC starting grant project LOCALSTAR. JEP thanks Exeter University for funding visits to the Exeter Astrophysics group.
Calculations were performed on facilities at Garching, and the University of Exeter Supercomputer, a DiRAC Facility jointly funded by STFC, the Large Facilities Capital Fund of BIS, and the University of Exeter.  Figure~1 was produced using \textsc{splash} \citep{splash2007}.

\bibliographystyle{mn2e}
\bibliography{Dobbs}

\bsp
\label{lastpage}
\end{document}